\documentclass[sigconf]{acmart}

\usepackage{booktabs}
\usepackage{multirow}
\usepackage{microtype}
\usepackage{xspace}
\usepackage{balance}
\acmConference[AgenticSE'26]{KDD 2026 Workshop on Agentic Software Engineering}{August 2026}{Jeju, Korea}
\acmYear{2026}
\acmISBN{}
\acmDOI{}

\title{Beyond Simpson's Paradox: A Cascade of Confounders in AI Agent Pull-Request Co-Authorship}

\author{Haoran Yu}
\affiliation{%
  \institution{Independent Researcher}
  \city{Seattle}
  \state{WA}
  \country{USA}}
\email{haoranyu889@gmail.com}

\author{Xiaochong Jiang}
\affiliation{%
  \institution{Independent Researcher}
  \city{Seattle}
  \state{WA}
  \country{USA}}
\email{jiang.xiaoc@northeastern.edu}

\author{Lifei Liu}
\affiliation{%
  \institution{Independent Researcher}
  \city{Seattle}
  \state{WA}
  \country{USA}}
\email{lliu.lifei@gmail.com}

\author{Su Wang}
\affiliation{%
  \institution{Carnegie Mellon University}
  \city{Pittsburgh}
  \state{PA}
  \country{USA}}
\email{suwang@alumni.cmu.edu}

\author{Pin Qian}
\affiliation{%
  \institution{Carnegie Mellon University}
  \city{Pittsburgh}
  \state{PA}
  \country{USA}}
\email{pqian@alumni.cmu.edu}

\author{Yihang Chen}
\affiliation{%
  \institution{Georgia Institute of Technology}
  \city{Atlanta}
  \state{GA}
  \country{USA}}
\email{ychen3726@gatech.edu}

\begin{abstract}
Pooled across five AI coding agents, pull requests (PRs) with a human
\texttt{Co-Authored-By} trailer merge \emph{less} often than purely-autonomous
ones (53.8\% vs.\ 79.8\%)---yet this aggregate finding is a textbook
\emph{Simpson's Paradox}.  Stratifying 33,596 PRs from the AIDev dataset by
agent identity reverses the conclusion: Copilot and Devin show large
positive within-agent gaps ($+41.2$ and $+33.5$\,pp, both $p{<}0.001$),
while Cursor, Claude Code, and Codex show small effects whose
cross-sectional 95\% CIs span zero.  The paradox is driven entirely by
agent composition: Codex, which dominates 64.9\% of the dataset, achieves
high merge rates while rarely using co-authorship.  But Simpson's Paradox
is only the first layer of a \emph{cascade of confounders}: within-repo
controls eliminate Devin's gap ($+33.5\!\to\!+1.6$\,pp, $p=0.73$); a
commit-count control further halves Copilot's within-repo gap
($+36.2\!\to\!+24.4$\,pp); restricted to multi-commit PRs, the Copilot
within-repo effect dissolves to $+4.8$\,pp ($p=0.59$).  No agent retains
a clear co-authorship effect once both repository selection and PR
structure are controlled.  Our findings caution against reporting
agent-pooled statistics without stratification and demonstrate that
cross-sectional co-authorship associations are largely selection and
PR-structure artefacts rather than evidence of a causal benefit.
\end{abstract}

\ccsdesc[500]{Software and its engineering~Software creation and management}
\ccsdesc[300]{Computing methodologies~Machine learning}

\keywords{AI coding agents, pull requests, Simpson's paradox, confounding,
co-authorship, human-AI collaboration, empirical software engineering}

\begin{document}
\maketitle

%% ================================================================
\section{Introduction}
%% ================================================================

AI coding agents (GitHub Copilot, Devin, Cursor, Codex, and
Claude Code), built on large language models trained on
code~\cite{chen2021codex}, can now draft, commit, and open pull requests (PRs)
with minimal human direction~\cite{xia2024agentless,dakhel2023github}.  They are part of a
broader proliferation of large-language-model-based systems across software and
data-intensive applications~\cite{cheng2026toward}.  A central question
for practitioners is: \emph{how often do these PRs actually get merged, and does
human involvement matter?}

A first-pass answer is deceptively simple to compute.  Using the
\texttt{Co-Authored-By} trailer that GitHub and many agent workflows insert into
commit messages~\cite{li2025aiteammates}, one can classify each PR as
\emph{human-agent collaborative} or \emph{purely autonomous}.  Pooling all
33,596 PRs from five agents, collaborative PRs merge at only 53.8\% while
autonomous PRs merge at 79.8\%, a gap of $-26.0$\,pp that appears to
condemn human participation as counterproductive.

We show this conclusion is entirely spurious.  The aggregate signal is driven by
\emph{agent composition}, not by co-authorship itself: Codex, which accounts for
64.9\% of all PRs and achieves high merge rates, rarely uses
\texttt{Co-Authored-By} (1.2\% of its PRs), while Copilot and Devin, whose
autonomous PRs succeed at 6.2\% and 21.6\% respectively, rely on
co-authorship for over 89\% and 95\% of their submissions.
Stratifying by agent, four of five agents show a positive
within-agent co-authorship effect, a classic reversal that constitutes
\textbf{Simpson's Paradox}~\cite{simpson1951interpretation,pearl2014comment}.

However, the within-agent reversal is itself the start of a longer story.
A staged robustness analysis exposes a \emph{cascade of confounders}:
within-repo controls collapse Devin's apparent gap to $+1.6$\,pp, and a
commit-count control further halves Copilot's within-repo effect, which
dissolves entirely ($+4.8$\,pp, $p{=}0.59$) when restricted to multi-commit
PRs.  No agent retains a clear within-stratum co-authorship effect once
both repository selection and PR structure are controlled.  Our data are
purely descriptive: causality cannot be established from observational
PR records.

This paper addresses three research questions:
\begin{description}
  \item[\textbf{RQ1.}] Does the presence of a \texttt{Co-Authored-By} trailer
    in agent-submitted PRs correlate with merge rate, and does this relationship
    hold within individual agents?
  \item[\textbf{RQ2.}] How do author/committer collaboration modes (fully
    autonomous, agent-draft, human-both) relate to PR merge outcomes?
  \item[\textbf{RQ3.}] Does adopting a second agent in a repository change
    existing merge rates?
\end{description}

%% ================================================================
\section{Background and Related Work}
\label{sec:related}
%% ================================================================

\paragraph{Pull-request dynamics.}
Gousios et al.~\cite{gousios2016work} established that reviewer workload,
PR size, and contributor reputation are strong predictors of merge outcomes.
Yu et al.~\cite{yu2016reviewer} showed that reviewer assignment strongly
influences acceptance.  We extend these lines of inquiry to the emerging setting
of AI-generated PRs.

\paragraph{Bots and automation in SE.}
Wessel et al.~\cite{wessel2020effects,wessel2021bots} found that adopting code
review bots on open-source projects changes PR workflows and contributor
behaviour; a follow-up study~\cite{wessel2022quality} confirmed downstream effects
on PR acceptance rates.  Our work complements this literature by studying
\emph{generative} AI agents, not merely notification/review bots.

\paragraph{AI agents and code generation.}
Dakhel et al.~\cite{dakhel2023github} evaluated Copilot's code quality;
Xia et al.~\cite{xia2024agentless} introduced Agentless, a minimal agent
baseline.  Neither study focuses on PR-level merge outcomes or human
collaboration signals, which is our contribution.

\paragraph{Simpson's Paradox.}
Simpson's Paradox~\cite{simpson1951interpretation} arises when a
confounding variable (here, \emph{agent identity}) drives opposite
associations at the aggregate and stratum levels.
Pearl~\cite{pearl2014comment} formalises conditions under which the
within-stratum analysis is causally preferred; we invoke this to argue that
the per-agent results are the correct lens for our data.

%% ================================================================
\section{Method}
\label{sec:method}
%% ================================================================

\paragraph{Dataset.}
We use the AIDev dataset~\cite{li2025aiteammates}, which contains 33,596 pull
requests submitted by five AI coding agents: Codex ($n=21,799$, 64.9\%),
Copilot ($n=4,970$, 14.8\%), Devin ($n=4,827$, 14.4\%), Cursor ($n=1,541$,
4.6\%), and Claude Code ($n=459$, 1.4\%).
PRs span 2024-12 to 2025-07 (218 days), but agent observation windows
differ: Devin appears earliest (2024-12), Codex latest (from 2025-05);
within-agent and within-repo analyses absorb time-invariant confounding,
but readers should bear this asymmetry in mind when interpreting pooled
counts.  Each PR is linked to its hosting repository and records the
final PR state (merged, closed-unmerged, or open at collection time).

\paragraph{Co-Authored-By mining.}
GitHub's commit-message convention supports a \texttt{Co-Authored-By: Name
<email>} trailer~\cite{li2025aiteammates}.  Agent workflows (e.g., Copilot
Workspace, Devin) routinely inject this trailer to attribute human editors.
We parse all commit messages in each PR and mark a PR as
\emph{collaborative} (\texttt{coauth=1}) if at least one commit contains a
\texttt{Co-Authored-By} trailer; otherwise the PR is marked
\emph{purely autonomous} (\texttt{coauth=0}).  In total, 10,764 PRs
(32.0\%) are collaborative and 22,832 (68.0\%) are purely autonomous.
Because the trailer is agent-inserted (most often by Copilot Workspace
or Devin to attribute human editors), in practice these flagged
co-authors are humans rather than bots; we discuss this assumption as
a measurement limitation in Section~\ref{sec:disc}.

\paragraph{Author/Committer classification.}
Beyond the trailer, GitHub exposes a distinction between the
\emph{author} (who wrote the commit) and the \emph{committer} (who applied
it to the branch).  We derive three collaboration modes:
(i)~\textbf{human\_both}: human author \emph{and} human committer;
(ii)~\textbf{agent\_draft}: bot author, human committer (human applied the
agent's commit); and
(iii)~\textbf{fully\_autonomous}: bot author \emph{and} bot committer.
A fourth logical combination (human author, bot committer) occurs in only
16 PRs and is excluded from the analysis due to insufficient sample size.

\paragraph{Outcome.}
The binary outcome is \emph{merge rate}: the proportion of PRs with
\texttt{state=merged} at the time of dataset collection.  We treat
PRs that are still open at collection time (2,312 PRs, 6.9\%) as
``not merged'' for conservatism, acknowledging this may undercount
eventual merges for recent PRs.  As a robustness check, restricting
to \texttt{state=closed} PRs (excluding the open ones) preserves the
qualitative findings: the pooled paradox direction holds ($-21.5$\,pp)
and Copilot's pure-autonomous merge rate remains low (7.1\% vs.\ the
6.2\% headline figure).

\paragraph{Identifying Simpson's Paradox.}
A Simpson's Paradox occurs when an aggregate association has the
\emph{opposite sign} to the within-stratum association for most or all
strata.  We operationalise this as: (1) pooled co-authorship effect
$<0$, and (2) majority of agents showing co-authorship effect $>0$.
We additionally verify that the effect reversal is explained by the
distributional imbalance in agent-level coauth rates (i.e.\ that
high-merge agents are systematically low-coauth).

\paragraph{Statistical tests.}
We use Pearson $\chi^2$ tests (with continuity correction) for proportion
comparisons and report two-sided $p$-values.  Effect sizes are reported
as percentage-point differences ($\Delta$\,pp).  For the multi-agent
analysis (RQ3) we use a difference-in-differences regression on weekly
merge rates, controlling for repository fixed effects and a linear time
trend.  No causal model is assumed; all relationships are
described as associations.

%% ================================================================
\section{Results}
\label{sec:results}
%% ================================================================

\subsection{RQ1: The Simpson's Paradox in Co-Authorship}
\label{sec:rq1}

\paragraph{Pooled analysis (misleading).}
Table~\ref{tab:pooled} shows the aggregate result.  Co-authored PRs merge at
53.8\% while purely-autonomous PRs merge at 79.8\%, a difference of
$-26.0$\,pp ($\chi^2=2424$, $p{\approx}0$).  A practitioner stopping here
would conclude that human co-authorship \emph{reduces} merge probability.

\begin{table}[t]
\centering
\caption{Pooled merge rates by co-authorship status (all agents combined).
The $-26.0$\,pp gap is entirely confounded by agent composition.}
\label{tab:pooled}
\begin{tabular}{lrrr}
\toprule
\textbf{Co-Authored-By} & \textbf{$n$ PRs} & \textbf{Merge Rate} & \\
\midrule
Yes (collaborative)   & 10,764 & 53.8\% & \\
No  (pure agent)      & 22,832 & 79.8\% & \\
\midrule
\textbf{Delta}        &        & $-26.0$\,pp & $\chi^2=2424$, $p{\approx}0$ \\
\bottomrule
\end{tabular}
\end{table}

\paragraph{Per-agent analysis (true picture).}
Table~\ref{tab:peragent} stratifies by agent identity.  The within-agent
picture almost completely reverses the pooled finding.

\begin{table}[t]
\centering
\caption{Per-agent merge rates by co-authorship status.
Copilot and Devin show large, highly significant positive gaps;
Codex's $-4.4$\,pp and the small positive gaps for Cursor and
Claude Code have 95\% CIs that include zero.
The pooled negative is an artefact of agent composition (Simpson's Paradox).}
\label{tab:peragent}
\small
\begin{tabular}{lrrrr}
\toprule
\textbf{Agent}
  & \textbf{\% coauth}
  & \textbf{MR (coauth)}
  & \textbf{MR (pure)}
  & \textbf{$\Delta$} \\
\midrule
Codex        &  1.2\% & 78.3\% & 82.6\% & $-4.4$\,pp \\
Copilot      & 89.4\% & 47.4\% &  6.2\% & $+41.2$\,pp \\
Cursor       & 69.8\% & 66.3\% & 62.7\% & $+3.7$\,pp  \\
Devin        & 95.9\% & 55.1\% & 21.6\% & $+33.5$\,pp \\
Claude Code  & 76.7\% & 60.2\% & 55.1\% & $+5.1$\,pp  \\
\bottomrule
\end{tabular}
\end{table}

\paragraph{Anatomy of the paradox.}
The confound is Codex: it contributes 64.9\% of all PRs, achieves an 82.6\%
merge rate, and uses \texttt{Co-Authored-By} in only 1.2\% of its submissions,
inflating the pooled autonomous rate far above any individual agent's.
Conversely, Copilot and Devin (both high-coauth) have much lower overall
merge rates, depressing the pooled collaborative rate.  The within-agent
gaps are striking: $+41.2$\,pp for Copilot and $+33.5$\,pp for Devin,
with pure-autonomous merge rates of just 6.2\% and 21.6\%.

Of Copilot's 528 purely-autonomous PRs, 54.5\% are explicit drafts
(\texttt{[WIP]} in title); excluding these raises pure-autonomous MR
from 6.2\% to 13.3\%.  Devin's 199 pure-autonomous PRs contain no
WIP markers.  By contrast, Codex's tiny $-4.4$\,pp gap is consistent
with its low co-authorship rate reflecting \emph{design choice} rather
than neglect.

\paragraph{Within-repo robustness check.}
To absorb repository-level confounders, we estimate
$\text{merged}_{i} = \alpha_{r,a} + \beta\cdot\text{coauth}_{i} + \varepsilon_{i}$
with repo$\times$agent pair fixed effects on the 391 repo--agent pairs
(12,301 PRs) with both co-authored and pure PRs (cluster-robust SEs at
pair level).  The overall effect is $+15.2$\,pp (SE${}=2.5$,
$p{<}0.001$, 95\% CI $[10.4, 20.0]$\,pp).

Crucially, this effect is driven almost entirely by \textbf{Copilot}
($+36.2$\,pp, $p{<}0.001$); a leave-one-out check excluding the top-3
Copilot repositories (microsoft/vscode, dotnet/aspire, mlflow/mlflow)
yields $+35.4$ to $+36.4$\,pp, so the result is not a single-repo artefact.
However, Copilot's pure-autonomous PRs are mostly single-commit
(median 1 vs.\ 4 for co-authored), and adding $\log(\text{commits})$ as
a control reduces the within-repo Copilot effect to $+24.4$\,pp
($p{<}10^{-10}$).  Restricting to multi-commit PRs (which removes the
single-commit-draft confound), the within-repo Copilot effect is
$+4.8$\,pp ($p=0.59$): \emph{not statistically distinguishable from zero}.
The headline Copilot association therefore reflects a structural workflow
correlate (single-commit drafts vs.\ multi-commit submissions) as much
as human co-authorship per se. The multi-commit Copilot null is also
underpowered: only $n{=}38$ multi-commit pure-autonomous Copilot PRs
exist, giving a minimum detectable effect of roughly $\pm 22$\,pp at
80\% power, so we cannot rule out moderate effects.
Devin's cross-sectional gap ($+33.5$\,pp) collapses to $+1.6$\,pp
($p=0.73$) within repos.
Codex ($+2.5$\,pp), Cursor ($+2.1$\,pp), and Claude Code ($+7.4$\,pp) also
show near-zero within-repo effects, though Codex's within-repo sample is
concentrated (top-3 repos contribute 69\% of its 6,186 PRs).  The practical
implication is more cautious than the cross-sectional numbers suggest:
once repository selection \emph{and} PR structure are controlled, no
agent shows a clear within-stratum co-authorship effect.

\subsection{RQ2: Author/Committer Collaboration Modes}
\label{sec:rq2}

Table~\ref{tab:modes} presents merge rates across the three collaboration
modes derived from author/committer attribution.

\begin{table}[t]
\centering
\caption{Merge rates by author/committer collaboration mode across all agents
($n=33{,}564$ PRs; excludes 16 human-author/bot-committer PRs and 16 PRs
lacking commit metadata).}
\label{tab:modes}
\begin{tabular}{lrr}
\toprule
\textbf{Mode}           & \textbf{$n$ PRs} & \textbf{Merge Rate} \\
\midrule
\texttt{human\_both}        & 23,200 & 82.0\% \\
\texttt{agent\_draft}       &    336 & 64.3\% \\
\texttt{fully\_autonomous}  & 10,028 & 47.6\% \\
\bottomrule
\end{tabular}
\end{table}

The pattern is strictly monotone: every additional step of human involvement
is associated with a higher merge rate.  Moving from fully-autonomous to
agent-drafted (human committer) raises the merge rate by $+16.7$\,pp; moving
further to human-both raises it by a further $+17.7$\,pp.  Crucially, this
gradient \emph{survives} the commit-count control that dissolved the RQ1
co-authorship effect: within every commit-count bin (1, 2, 3--5, $>$5
commits) and within every agent with sufficient data, \texttt{human\_both}
exceeds \texttt{fully\_autonomous} by 25--49\,pp.  The author/committer
signal, unlike the \texttt{Co-Authored-By} marker, is not absorbed by
PR-structure controls.

The small \texttt{agent\_draft} category (336 PRs) at 64.3\% sits in
an informative intermediate position, consistent with human sign-off
compensating for agent-authored content.

\subsection{RQ3: Multi-Agent Adoption Patterns}
\label{sec:rq3}

We briefly examined whether adopting a second agent changes existing merge rates
(233 multi-agent repos).  A difference-in-differences regression on weekly merge
rates, controlling for repository fixed effects and a linear time trend, estimates
a treatment effect of $-12.1$\,pp (SE${}=3.6$, $p{<}0.001$, 95\% CI
$[-19.2, -5.0]$\,pp).  This suggests that introducing a second agent is
associated with a meaningful decline in merge rates, though the effect may
reflect compositional changes (the second agent contributing lower-quality PRs)
rather than disruption to the first agent's workflow.

%% ================================================================
\section{Discussion and Limitations}
\label{sec:disc}
%% ================================================================

\paragraph{Direction of causality.}
Our data are purely observational.  The positive within-agent
association between co-authorship and merge rate does \emph{not} imply that
adding a human co-author to a PR \emph{causes} it to be merged.  An equally
plausible interpretation is \emph{selection}: developers may invest effort in
co-authoring only those agent PRs they already believe are worthwhile, so
\texttt{Co-Authored-By} marks quality rather than creating it.
The cascade of confounders documented in \S\ref{sec:rq1} is consistent
with this: pure-autonomous Copilot PRs are mostly single-commit drafts
(54.5\% \texttt{[WIP]}), so the cross-sectional gap captures PR
maturation, repository selection, and curation as much as co-authorship
\emph{per se}. Full causal identification would require a randomised
experiment.

\paragraph{Agent composition dominance.}
The Simpson's Paradox is manufactured almost entirely by Codex, which contributes
64.9\% of the dataset.  Interpretations of pooled statistics in multi-agent
datasets must control for agent identity, just as cross-study meta-analyses must
control for study characteristics.  We recommend against reporting agent-pooled
merge rates without stratification in future empirical work on AI agents.

\paragraph{Measurement limitations.}
The \texttt{Co-Authored-By} trailer is agent-inserted and not standardised;
some agents may omit it even when humans edited commits.  Our headline
definition counts any trailer as ``collaborative,'' which is conservative
for Copilot/Cursor/Devin (where $\geq 99\%$ of trailers are human emails)
but inflates apparent collaboration for Claude Code and Codex, which often
self-attribute via \texttt{noreply@anthropic.com}.  Restricting to non-bot
co-author emails leaves the within-repo Copilot finding essentially
unchanged ($+36.1$ vs $+36.2$\,pp) but reveals a previously masked
positive Claude Code effect (cross-sectional $+33.8$\,pp, $n{=}47$).
We measure merge rate at collection time; open PRs are coded as ``not merged.''

\paragraph{Scope.}
This study covers five agents in AIDev and may not generalise beyond them.
We do not control for PR size, task type, or reviewer
experience~\cite{gousios2016work,yu2016reviewer}, and the dataset
exhibits survivorship bias (only repos that adopted agents are included).

\paragraph{A collaboration-level framework.}
Our results motivate a four-level taxonomy of human involvement in
agent-generated PRs, ordered by increasing human participation:

\begin{table}[t]
\centering
\caption{A collaboration-level framework for human involvement in
agent-generated PRs, ordered by increasing human participation.
L0/L1/L3 merge rates from author/committer classification (Table~\ref{tab:modes});
L2 from co-authorship analysis (pooled).}
\label{tab:levels}
\small
\begin{tabular}{@{}llr@{}}
\toprule
\textbf{Level} & \textbf{Description} & \textbf{MR} \\
\midrule
L0: Fully autonomous & bot author + bot committer & 47.6\% \\
L1: Agent draft      & bot author, human committer & 64.3\% \\
L2: Co-authored      & \texttt{Co-Authored-By} present & $^*$ \\
L3: Human-both       & human author + human committer & 82.0\% \\
\bottomrule
\multicolumn{3}{@{}l}{\footnotesize $^*$Pooled 53.8\%; within-repo $+15.2$\,pp.}
\end{tabular}
\end{table}

\noindent Combining this with the within-repo analysis yields
agent-specific deployment recommendations:

\begin{table}[t]
\centering
\caption{Agent-specific deployment recommendations. The ``Min.\ Level'' column
indicates the minimum collaboration level at which the within-repo evidence
supports deployment.}
\label{tab:recommendations}
\small
\begin{tabular}{@{}llr@{}}
\toprule
\textbf{Agent} & \textbf{Min.\ Level} & \textbf{Within-repo $\Delta$} \\
\midrule
Codex       & L0 (autonomous OK) & $+2.5$\,pp ($p{=}0.40$) \\
Copilot     & L2+ tentative$^\dagger$ & $+36.2$\,pp ($p{<}0.001$) \\
Cursor      & L0 (autonomous OK) & $+2.1$\,pp ($p{=}0.62$) \\
Devin       & L0 (autonomous OK) & $+1.6$\,pp ($p{=}0.73$) \\
Claude Code & L0--L1 (tentative) & $+7.4$\,pp ($p{=}0.35$) \\
\bottomrule
\multicolumn{3}{@{}l}{\footnotesize $^\dagger$Within-repo Copilot effect drops to $+4.8$\,pp ($p{=}0.59$)}\\
\multicolumn{3}{@{}l}{\footnotesize when restricted to multi-commit PRs.}
\end{tabular}
\end{table}

As Table~\ref{tab:recommendations} shows, Copilot is the only agent with
a significant within-repo co-authorship association at the cross-sectional
level, but this effect is largely explained by PR structure: restricted
to multi-commit PRs, Copilot's within-repo effect is $+4.8$\,pp
($p{=}0.59$).  For Codex, Cursor, and Devin, autonomous operation
shows no within-repo penalty.  Practically, teams deploying Copilot
should be aware that the apparent benefit of mandating co-authorship is
weaker than headline numbers suggest once PR structure is controlled;
the strongest practical signal remains Copilot's overall low
pure-autonomous merge rate ($\sim$13\% non-WIP), which itself reflects
that pure-autonomous Copilot PRs are predominantly single-commit drafts.

%% ================================================================
\section{Conclusion}
\label{sec:conclusion}
%% ================================================================

We documented a textbook Simpson's Paradox in AI agent pull-request data:
pooled across five agents, co-authorship appears to \emph{hurt} merge rates
($-26.0$\,pp), yet within-agent analysis reveals positive gaps for four of
five agents, with Copilot and Devin showing $+41.2$ and $+33.5$\,pp
cross-sectionally.  This reversal, however, is only the first layer of a
\emph{cascade of confounders}: within-repo controls collapse Devin's gap
to $+1.6$\,pp, and a commit-count control reduces Copilot's within-repo
gap from $+36.2$ to $+24.4$\,pp; restricted to multi-commit PRs the
Copilot effect is $+4.8$\,pp ($p{=}0.59$).  A complementary
author/committer analysis confirms a graded pattern
(47.6\% $\to$ 64.3\% $\to$ 82.0\%).

These findings argue for agent-stratified analyses in SE empirical studies
and demonstrate that cross-sectional co-authorship associations can be
misleading without within-repo \emph{and} within-PR-structure controls:
what appears as a universal or even agent-specific human-involvement
benefit is, once both confounders are controlled, more cautiously
characterised as a structural correlate (single-commit drafts vs.\
multi-commit submissions) than evidence of a causal benefit of
co-authorship per se.
Analysis scripts are available for replication at
\url{https://anonymous.4open.science/r/pr-coauth-simpsons-paradox-7DC2}.

%% ================================================================
\begin{acks}
We thank the maintainers of the AIDev dataset for making this research
possible, and the anonymous reviewers for their constructive feedback.
\end{acks}

\bibliographystyle{ACM-Reference-Format}
\balance
\bibliography{references}

\end{document}